\documentclass[onecolumn,preprintnumbers,amsmath,amssymb]{revtex4}

\usepackage{graphicx}
\usepackage[latin1]{inputenc}
\usepackage{dcolumn}
\usepackage{bm}
\usepackage{epsfig}
\usepackage[english]{babel}
\usepackage{blindtext}
\usepackage{lipsum}
\usepackage{color}

\usepackage{floatrow}
\newfloatcommand{capbtabbox}{table}[][\FBwidth]
\begin{document}


\title{\textsf{Election Forensics: quantitative methods for electoral fraud detection}}

\author{Lucas Lacasa} 
\email{l.lacasa@qmul.ac.uk}
\affiliation{School of Mathematical Sciences, Queen Mary University of London, Mile End Road, London E14NS (UK)}%
\author{Juan Fern\'andez-Gracia} 
\email{juanf@ifisc.uib-csic.es}
\affiliation{Instituto de Fisica Interdisciplinar y Sistemas Complejos (IFISC), CSIC-UIB, Palma de Mallorca E-07122 (Spain)}



 \maketitle


The last decade has witnessed an explosion on the computational power and a parallel increase of the access to large sets of data --the so called Big Data paradigm-- which is enabling to develop brand new quantitative strategies underpinning description, understanding and control of complex scenarios. One interesting area of application concerns fraud detection from online data, and more particularly extracting meaningful information from massive digital fingerprints of electoral activity to detect, a posteriori, evidence of fraudulent behavior. 
In this short article we discuss a few {\it quantitative} methodologies that have emerged in recent years on this respect, which altogether form the nascent interdisciplinary field of election forensics. Aiming to foster discussion and raise awareness on this interdisciplinary area, we hereby enumerate a few of the most relevant approaches and methods.\\ 

\noindent {\bf \textsf{Benford's law. }}The first example of these methodologies evaluates the deviation or conformance of vote counts statistics to the so-called {\it Benford's law} \cite{benford, hill2, nigrini, mebane}, that predicts that the first significant digits in some datasets (including vote counts) should follow a decaying logarithmic distribution.
The first significant digit (or leading digit) of a number is defined as its non-zero leftmost digit (for instance the leading digit of 123 is 1 whereas the leading digit of 0.025 is 2). Benford's law is an empirical statistical law stating that in particular types of numeric datasets the probability of finding an entry whose first significant digit is $d$ decays logarithmically as
\begin{equation}
P(d)=\log_{10}(1+1/d),
\label{BL}
\end{equation}
where $\log_{10}$ stands here for the decimal logarithm (note that trivially $\sum_{d=1}^9 P(d)=1)$. Perhaps counterintuitively, this law is quite different from the expected distribution arising from an uncorrelated random process (e.g. coin tossing or extracting numbers at random from an urn) which would yield a uniform distribution where every leading digit would be equally likely to appear.
The shape given in Eq.\ref{BL} was empirically found first in 1881 by astronomer Simon Newcomb and later popularized and exhaustively studied by Frank Benford \cite{benford}. Empirical datasets that comply to Benford's law emerge in as disparate places as for stock prices or physical constants, and some mathematical sequences such as binomial arrays or some geometric sequences have been shown to conform to Benford. A possible origin of this law has been rigorously explained by Hill \cite{hill}, who proved a central limit-type theorem by which random entries picked from random distributions form a sequence whose leading digit distribution converges to Benford's law. Another explanation comes from the theory of multiplicative processes, as it is well known that power-law distributed stochastic processes follow Benford's law for the specific case of a density $1/x$ (see \cite{tolo} and references therein for details). In practice, this law is expected to emerge in a range of empirical datasets where part or all of the following criteria hold: (i) the data ranges a broad interval encompassing several orders of magnitude rather uniformly, (ii) the data are the outcome of different random processes with different probability densities, (iii) the data are the result of one or several multiplicative processes. \\

\begin{figure}
\begin{floatrow}
\capbtabbox{%
 \begin{tabular}{|c|c|c|c|}
 \hline
  \textbf{Constituency} & \textbf{Municipality} & \textbf{Party} & \textbf{Vote count} \\ \hline 
A Coru\~na & A Ba\~na & PP & 1170 \\
A Coru\~na & A Ba\~na & PSOE & 678 \\
A Coru\~na & A Ba\~na & PODEMOS & 325 \\
A Coru\~na & A Ba\~na & C's & 139 \\
A Coru\~na & A Capela & PSOE & 324 \\
A Coru\~na & A Capela & PP & 192 \\
A Coru\~na & A Capela & PODEMOS & 181 \\
A Coru\~na & A Capela & C's & 42 \\
\vdots & \vdots & \vdots & \vdots\\
\hline
 \end{tabular}
}{%
  \caption{Sample of Spanish election results of 2015. The results are shown for the four biggest parties in two different municipalities that belong to the same electoral district (constituency). The first two columns aggregates data in different spatial scales, the third column displays the party that is investigated while the fourth column provides the vote counts. For 1BL only the leading digit is of interest, while for 2BL only the second digit is.}\label{tab:results}%
}
\ffigbox{%
   \includegraphics[width=0.4\textwidth]{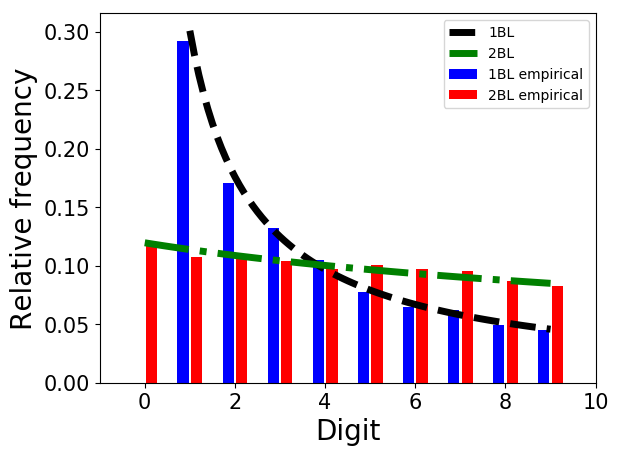}%
}{%
  \caption{Benford's laws (1BL and 2BL). The bars correspond to 1BL (blue) and 2BL (red) at the level of municipalities for PP in the Spanish 2015 national elections. The dashed black line corresponds to the theoretical 1BL values and the dashed-dotted to 2BL. From this plot one can compute the distance between theory and empirical data with the use of a $\chi^2$ test or other statistical tests in order to accept/discard the hypothesis that the data follows 1BL/2BL. } \label{fig:BL}%
}
\end{floatrow}
\end{figure}

\noindent Advocated by some academics including Nigrini \cite{nigrini}, the application of Benford's law to detect fraud and irregularities --by observing anomalous and statistically significant deviations from Eq.\ref{BL} for datasets which otherwise should conform to that distribution-- has become popular in recent years, and from now on we quote this a 1BL test. While some authors have tested 1BL in some electoral data \cite{mansilla, roukema}, other authors \cite{mebane, Bayesian} strongly advocate instead to look at the {\it second} significant digit (which follows an extended version of Benford's law \cite{hill2}) and argue that the frequencies of election vote counts at precinct level approximate a Benford distribution for the second digit, and accordingly mismanaged or fraudulent manipulation of vote counts would induce a statistically significant deviation in the distribution of the second leading digit, detected by a simple Pearson $\chi^2$ goodness of fit test. Mebane \cite{mebane} applied this so-called 2BL test to assess the cases of Florida 2004 and Mexico 2006, and other authors have subsequently applied this in many other occasions (see \cite{juan,germany,Bayesian} and references therein). In this case the theoretical distribution takes a more convoluted shape than Eq.\ref{BL}, namely
\begin{equation}
P_2(d)=\sum_{k=1}^{9}\log_{10}\left(1+\frac{1}{10k+d}\right),
\label{2BL}
\end{equation}
and a good numerical approximation \cite{hill2,germany} is given by
$$P_2(d)\simeq(0.11968, 0.11389, 0.10882, 0.10433, 0.10031, 0.09668, 0.09337, 0.09035, 0.08757, 0.08500).$$
In order to quantitatively conclude whether an empirical distribution is compatible with the expected (theoretical) one, it is customary to resort to statistical conformance tests. A classical option is to use the well-known Pearson's $\chi^2$ statistic
$$\chi^2=N\sum_{d=m}^9\frac{[P_{\text{obs}}(d)-P_{\text{th}}(d)]^2}{P_{\text{th}}(d)},$$
where $P_{\text{th}}(d)$ and $P_{\text{obs}}(d)$ are the theoretical and observed relative frequencies of each digit, $N$ is the total sample size, and $m=1$ for 1BL and $m=0$ for 2BL. The null hypothesis  is that data conform to Benford's law. This statistic has 8 degrees of freedom for 1BL and 9 for 2BL (as in this latter case the digit zero has to be incorporated as a candidate) and is to be compared to certain critical values, such that if $\chi^2>\chi^2_{n,a}$ then the null hypothesis is rejected with the selected level of confidence level $a$. For $n=8$ degrees of freedom, the critical values at the $95\%$ and $99\%$ are 15.507 and 20.090 respectively, whereas for $n=9$ degrees of freedom the critical values at the $95\%$ and $99\%$ are 16.919 and 21.666 respectively.\\
Since Pearson's $\chi^2$ is known to sometimes suffer from being over rejected for large samples, alternative proposals have recently been put forward, such as Bayesian approaches for testing conformance \cite{Bayesian}.\\
In any case, the standard procedure considers vote count data (at possibly different possible resolution levels), computes histograms of 1BL and/or 2BL, and then performs statistical tests to conclude whether there is conformance to Benford's laws (see figure \ref{fig:BL} and table \ref{tab:results} for an illustration). As a word of caution, note that this methodology is not free of controversy \cite{contro1, contro2}. On the other hand, vote counts could be manipulated in such a way that BL holds (probably this is easier for 1BL than 2BL, for instance). In this sense, the usefulness of this approach resides on the premise that fraudsters are not aware of this empirical regularity emerging naturally in vote statistics, and the fact that carelessly manipulated data deviate from this law.  Also, evidence of deviation from BLs  might be due either to unintentional mismanagement of the voting process and/or to fraud. This type of analysis only flags the existence of such irregularities and gives no judgment on what was the cause for such irregularity. \\

\noindent {\bf \textsf{Integer percentages. }} Another rather simple but unexpectedly present statistical regularity in conceivably fraudulent election data is the overrepresentation of integer or round percentages of voteshare (percentage of votes to each party) and turnout (percentage of the total population that actually voted) \cite{new}. The reason for this is purely psychological: there exist scientific evidence that supports the fact that humans have a natural tendency to select round numbers or multiples of five (e.g. 50\%, 70\%, 75\%, etc), and then if both voteshare and turnout percentages are going to be made up `by hand', these psychological biases could be reflected. Similarly, it is more common/psychologically more straightforward to fabricate a percentage such as 91\% than 91.273\%, say. Note that percentages do not show up in raw, official vote count data, only when data are aggregated, so artificial overrepresentation of these is typically hidden. Overrepresenting these can be easily seen as periodic tiling patterns emerging in the co-occurrence heat maps \cite{new2} (see below).\\
Similarly, other sorts of psychological biases when it comes to making up election data have been exploited as well \cite{integer2}.\\

\begin{figure}
 \includegraphics[width=0.4\textwidth]{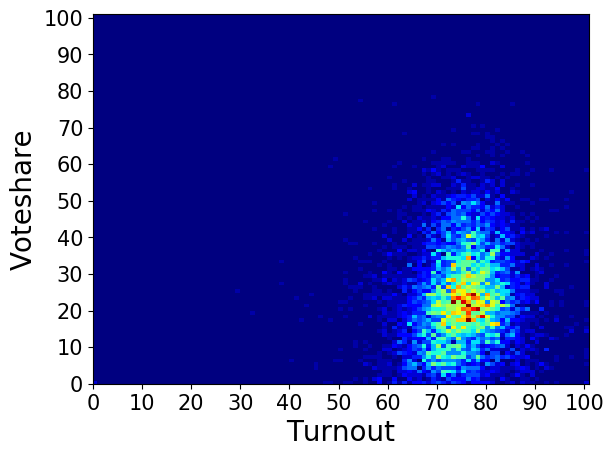}
 \includegraphics[width=0.4\textwidth]{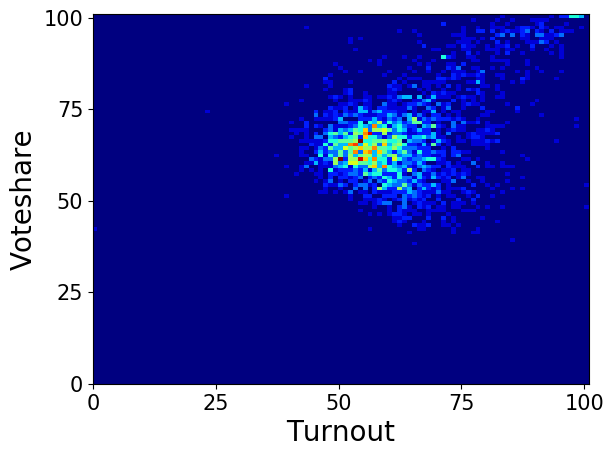}
 \caption{Co-occurrence heat maps. \textbf{Left:} Spanish socialist party results for 2015, with no apparent signs of fraudulent behavior. \textbf{Right:} Results of United Russia in the presidential elections of 2012. Note in this case how the data is skewed towards the corner representing 100\% turnout and 100\% voteshare.}
 \label{fig:co}
\end{figure}

\noindent {\bf \textsf{Co-occurence heat maps. }}
Another methodology for electoral forensic analysis relates to the presence and detection of sources of incremental and extreme fraud from the co-occurring statistics of vote and turnout numbers.
This analysis is inspired by a simple remark made by Sobyanin and Sukhovolsky \cite{new2}: suppose that the fraudulent mechanism is based on adding to the ballot box of a particular polling station a large number of fake ballots for some party (so-called ballot stuffing). In this case, not only the results of the favored party will improve there, but also the turnout for that polling station will increase, hence if that fraud strategy is systematically repeated one should expect a {\it positive correlation} between results of the favored party and turnout to emerge. Note that if instead of adding fake ballots the strategy relies on removing from the ballot box some genuine ballots voting for any other party from the one favored, then what emerges is a negative correlation between voteshare to the fraudulent party and total turnout in the polling station.\\
Additionally, a recent study \cite{klimek} capitalizes on this effect and explores the co-occurring statistics of vote and turnout numbers and the associated double mechanism of incremental and extreme fraud by plotting two dimensional histograms (heat maps) reporting, for a given political party, the percentage of vote (voteshare) it got as compared to the turnover. According to Klimek and co-authors, incremental fraud occurs when with a given rate, ballots for one party are added to the urn and/or votes for other parties are taken away (i.e., high correlation between voteshare and turnout). On the other hand, extreme fraud (which corresponds to reporting close to complete turnout and almost all votes for a single party) emerges when the distribution transitions {\it from unimodal to bimodal} and one of the modes corresponding to a cluster that concentrates close to that corner of $100 \%$ participation (complete turnout) and very large voteshare. An example with real data is depicted in figure \ref{fig:co}.\\
Note that this particular approach is concerned with vote rigging (which is the act of fraudulent alteration of the content of ballot boxes) which is different from {\it voter rigging}, i.e.  the unlawful and systematic harassment of the voters themselves, which can also be quantitatively detected as analyzed recently \cite{voter}.\\

\noindent To conclude, in recent years the field of {\it Election Forensics} --agglutinating quantitative approaches for a posteriori fraud detection in electoral data-- has emerged, providing compelling scientific evidence for flagging fraud in cases such as Russia elections \cite{new2}, Venezuela's post-recall referendum elections \cite{Venezuela}, or the recent Turkish constitutional referendum \cite{Turkish} which potentially paves the way from democracy to autocracy. In this short piece we have briefly discussed only a subset of recent quantitative methods for fraud detection.
Most of these approaches rely in one way or another on the presence or lack of {\it expected patterns}, i.e. statistical regularities which should be or not be present in fair election data. Probably the pros of these approaches rely on their simplicity (easy to implement) and interpretability. However, as already stated, it's also true that fraudsters aware of these regularities can try to make up data in such a way that specific patterns which emerge in fair data also hold even in the manipulated data. Then, what? Statistical and machine learning techniques could give us a hand here, by providing tools and algorithms that look for abstract patterns in the data. For instance, by using synthetic data of vote counts as well as empirical data from controlled scenarios (e.g. where fraud has been proved) one can feed algorithmic classifiers to `learn' the specific --yet non interpretable-- regularities present in the data and associate them with the existence or non-existence of fraudulent manipulation \cite{ML, ICA}.\\

\noindent All in all, election forensics is an emerging field which will probably be of utmost importance for an adequate and prolific implementation of new democratic tools such as electronic elections, referendums and other sorts of online polls.

%

\bibliography{apssamp}

\end{document}